# After antimatter gravity experiment at CERN - Prospects of the theory of the gravitational polarization of the quantum vacuum

Dragan Slavkov Hajdukovic

INFI, Cetinje, Montenegro

dragan@hajdukovic.com, dragan.hajdukovic@alumni.cern

**Abstract**

We start with a summary of the theory of the gravitational polarization of the quantum vacuum, *which makes it possible to consider the quantum vacuum as a source of gravity in the Universe*. Next, because virtual gravitational dipoles are indispensable for this theory, we suggested different theoretical possibilities for their existence. After that, we consider the first indications that three potential phenomena which, if confirmed, will present a major challenge to the Standard ΛCDM Cosmology, can be explained within the framework of the gravitational polarization of the quantum vacuum; this adds to the already existing successes of the theory. These potential phenomena include: non-Newtonian gravity in internal dynamics of Wide Binary Stars; the existence of massive relic galaxies, as NGC 1277, which is dark matter deficient; a surprisingly fast formation of the first stars and galaxies. Finally, as an example of the complexity of gravitational polarization, we demonstrated the existence of regions with a negative effective gravitational charge; the insufficient understanding of complexity of polarization was reason for premature claims that observations within the Solar System dismiss the gravitational polarization of the quantum vacuum. All this suggests that recent antimatter gravity experiment at CERN (which proved that atoms and antiatoms have gravitational charge of the same sign and hence ruled out the existence of a particular kind of gravitational dipoles) cannot be considered as a disproof of the theory. The existence of virtual gravitational dipoles and gravitational polarization of the quantum vacuum remain open questions and exciting possibilities for a new model of the Universe.

## 1. Introduction

*Is the quantum vacuum a source of gravity in the Universe*? No one knows the answer to this absolutely fundamental question; perhaps the most fundamental question within contemporary physics. Our scientific description of the Universe will be quite different if the quantum vacuum is a source of gravity, and consequently a cosmological fluid that must be included in the cosmological field equations.

The paradigm considered in this paper is that *the quantum vacuum is* (a so far completely neglected) *source of gravity in the Universe.* The inherent part of our paradigm is that *the quantum vacuum is a source of gravity through the mechanism of the gravitational polarisation that is caused by the immersed ordinary matter* (baryonic matter in astronomical jargon); such polarisation is possible if *at least some of the quantum vacuum fluctuations are virtual gravitational dipoles that can be aligned with the external gravitational field*. Of course, if we are not aware of a source of gravity, especially of an omnipresent source of gravity as quantum vacuum might be, phenomena caused by the disregarded source would be a surprise and mystery for us. If we remain completely unaware of the real source of gravity, we will eventually invent artificial theoretical constructions that can temporarily compensate our ignorance. It might be that dark matter and dark energy are such artificial theoretical constructs which, to some extent compensate for our disregard of the quantum vacuum as a source of gravity.

According to the Standard Model of Particles and Fields (which is together with General Relativity one of two cornerstones of contemporary physics), everything in the Universe (i.e. matter-energy content of the Universe) is made from apparently structureless, point-like particles (quarks and leptons) which interact through the exchange of gauge bosons (gluons for strong interactions; photons for electromagnetic interactions; $Z^0$, $W^+$ and $W^-$ bosons for weak interactions). This well-established content of the Universe can be called "matter of the Standard Model" (or Standard Model matter) and must be clearly distinguished from the non-established, i.e. the assumed content of the Universe, as for instance dark matter and dark energy.

In simple words, the quantum vacuum is one particular state of matter–energy of the Standard Model matter, radically different from more familiar states (solid, fluid, gas, ordinary plasma, quark–gluon plasma . . .) but as real as they are. You may envision quantum vacuum as an omnipresent fluid composed of quantum vacuum fluctuations, or, in a more popular wording, composed of virtual particle–antiparticle pairs with an extremely short lifetime (for instance, the lifetime of a virtual electron–positron pair (which is in fact a virtual electric dipole) should be of the order of $10^{-21}$ seconds). Note that within the Standard Model of Particles and Fields, quantum vacuum is the only state with the same amount of matter and antimatter.

We have empirical evidence that quantum vacuum and matter immersed in it interact through electromagnetic, strong and weak interactions (a brief description for non-physicists is given in Hajdukovic 2020 and a deeper introduction in Aitchison 2009); however, our knowledge of the gravitational properties of the quantum vacuum is at absolute zero. Worse than that, there is no empirical evidence suggesting in which direction to think about a possible gravitational impact of the quantum vacuum.

In principle, according to the Standard Model of Particles and Fields there are two possible sources of gravity; the first source is the quantum vacuum, while the second source is ordinary matter of the Standard Model, which is immersed in the quantum vacuum. From the cosmological constant problem (Weinberg, 1989) we know that these two sources of gravity must be radically different. Let us just remember the essence of the cosmological constant problem. If, using our theory of gravity, we calculate both the gravitational field caused by the quantum vacuum and the gravitational field caused by the immersed matter (stars, galaxies…), the gravitational field caused by the quantum vacuum is a minimum 40 orders of magnitude larger.

Consequently, if we want to consider quantum vacuum as a source of gravity, the gravitational field caused by the quantum vacuum must be many orders of magnitude weaker than predicted by our best theory of gravity. The simplest possibility to reduce the enormous gravitational field of the quantum vacuum is to assume that quantum vacuum fluctuations are virtual gravitational dipoles.

Section 2 of this paper is a summary of the theory of the gravitational polarization of the quantum vacuum (Hajdukovic 2011, 2014, 2020, 2021 and references therein), which makes it possible to consider the quantum vacuum as a source of gravity in the Universe. In Section 3 of this paper, faced with the fact that the existence of virtual gravitational dipoles is indispensable for the existence of gravitational polarisation, we point to very different theoretical possibilities for the existence of gravitational dipoles. Section 4 is devoted to the first indications that, three potential astronomical phenomena that emerged this year, can be explained within the framework of the gravitational polarization of the quantum vacuum; hence the hypothesis of the existence of virtual gravitational dipoles continues to be in apparent and encouraging agreement with astronomical observations and measurements. These three potential phenomena include: an apparent breakdown of the Newton–Einstein gravity in internal dynamics of Wide Binary Stars (Chae 2023; Hernandez 2023); the existence of massive relic galaxies, as NGC 1277 galaxy, which is dark matter deficient (Comerón 2023); a surprisingly fast formation of the first stars and galaxies (Ferreira et al.2023; Labbé et al. 2023). Section 5 is about two topics. First, as one example of the complexity of the gravitational polarization of the quantum vacuum, we focus on the regions with a negative effective gravitational charge. Second, we address the premature claims (Banik&Kroupa 2020) that the gravitational polarization of the quantum vacuum is dismissed by observations within the Solar System. Finally an Outlook is given in Section 6.

## 2. Summary of the theory of the gravitational polarization of the quantum vacuum

The working hypothesis that *quantum vacuum fluctuations are virtual gravitational dipoles* has two conflicting aspects (one unfavourable and one favourable) which open two different (complementary and challenging) fields of research.

The first, unfavourable aspect is that, according to our currently best knowledge, the existence of virtual gravitational dipoles is apparently unlikely. Because of this, nearly all physicists, without any consideration, immediately reject the theory as wrong.

The second (favourable) aspect is that, according to the pioneering initial studies (Hajdukovic 2011, 2014, 2020, 2021 and references therein), the assumption of the existence of virtual gravitational dipoles has consequences which are in good agreement with astronomical observations. If the hypothesis is wrong, the question is why the consequences are apparently correct.

The first field of research opened by our hypothesis is the experimental search for gravitational dipoles; comparable with the quest for magnetic monopoles. In fact the first human search for gravitational dipoles was recently performed by ALPHA-g collaboration, which demonstrated Anderson E.K. *et al.* 2023) that atoms of hydrogen and antihydrogen fall in the same direction in the gravitational field of the Earth, (in other words atoms and antiatoms have gravitational charge of the same sign). Consequently, while a proton-antiproton pair is an electric dipole composed of two electric charges of the opposite sign, it cannot be considered as a gravitational dipole composed of two gravitational charges of the opposite sign. This experiment shouldn't be considered as the end, but as the beginning of the experimental and astronomical search for gravitational dipoles, because in principle *different kinds of gravitational dipoles might exist* as we will show in the next section.

The second field of research is to study the astronomical and cosmological consequences of the assumed existence of virtual gravitational dipoles with an appropriate gravitational dipole moment. The already mentioned preliminary studies in this field of research have shown that the quantum vacuum enriched with virtual gravitational dipoles can be considered as a source of gravity causing phenomena which are in astonishing agreement with astronomical and cosmological observations. This is for the first time, that in an apparently successful way, the quantum vacuum was included as a major source of the gravitational field in the Universe.

The basis of the gravitational polarization of the quantum vacuum is simple and elegant.

If gravitational dipoles (with a non-zero gravitational dipole moment $p_g$) exist, the gravitational polarization density $P_g$, i.e. the gravitational dipole moment per unit volume, can be attributed to the quantum vacuum. It is obvious that the magnitude of the gravitational polarization density $P_g$ satisfies the inequality $0 \leq P_g \leq P_{gmax}$, where 0 corresponds to random orientations of dipoles, while the maximal magnitude $P_{gmax}$ corresponds to the case of saturation (when all dipoles are aligned with the external field). The value $P_{gmax}$ must be a universal constant related to the gravitational properties of the quantum vacuum.

If the external gravitational field is zero, quantum vacuum may be considered as a fluid of randomly oriented gravitational dipoles. In this case, the gravitational polarization density is equal to zero ($P_g \equiv 0$); of course, such a vacuum is not a source of gravitation.

However, the random orientation of virtual dipoles can be broken by the gravitational field of the immersed Standard Model matter. Massive bodies (particles, stars, planets, black holes, . . . ) but also many-body systems such as galaxies are surrounded by an invisible halo of gravitationally polarized quantum vacuum, i.e. a region of non-random orientation of virtual gravitational dipoles ($\boldsymbol{P}_g \neq 0$).

The magic of non-random orientation of dipoles, i.e. the magic of gravitational polarization of the quantum vacuum, is that the otherwise gravitationally featureless quantum vacuum becomes a source of gravity! And, let's repeat, a major source of gravity in the Universe.

## 2.1 The fundamental equations of the theory

This qualitative picture presented above, can be mathematically described in the following way.

The *spatial variation* of the gravitational polarization density generates *a gravitational bound charge density* of the quantum vacuum:

$$\rho_{qv} = -\boldsymbol{\nabla} \cdot \boldsymbol{P}_g \tag{1}$$

This gravitational bound charge density is in fact an effective gravitational charge density, which acts as if there is a real non-zero gravitational charge. That is how the magic of polarization works; quantum vacuum is a source of gravity, thanks to the immersed Standard Model matter.

In general, the magnitude $P_g$ is a function of the coordinates $(x_1, x_2, x_3)$, while the gravitational polarization density $\boldsymbol{P}_g$ and the corresponding Newtonian acceleration $\boldsymbol{g}_N$ (determined by distribution of the Standard model matter) point in the same direction. Consequently, we can write:

$$P_g = P_{gmax} f_g(x_1, x_2, x_3) \text{ and } \boldsymbol{P}_g = P_{gmax} f_g(x_1, x_2, x_3) \frac{\boldsymbol{g}_N}{g_N} \tag{2}$$

The values of function $f_g(x_1, x_2, x_3)$ belong to the interval [0, 1] where values 0 and 1 correspond, respectively, to the cases of random orientation of dipoles and saturation.

Equations (1) and (2) lead to

$$\rho_{qv}(x_1, x_2, x_3) = -P_{gmax} \boldsymbol{\nabla} \cdot \left[ f_g(x_1, x_2, x_3) \frac{\boldsymbol{g}_N}{g_N} \right] \tag{3}$$

Hence, in order to calculate the effective gravitational charge density of the quantum vacuum (analytically for N ≤ 2 and numerically for N ≥ 3 point-like bodies), we must know the function $f_g(x_1, x_2, x_3)$ and the unit vector $\boldsymbol{g}_N/g_N$ of the Newtonian acceleration caused by the distribution of the standard model matter.

It is important to understand the following. In the case of an ideal isolated system, $\boldsymbol{g}_N$ is the Newtonian gravitational field caused by the distribution of the Standard Model matter within the system. However, if the system is not isolated, $\boldsymbol{g}_N$ is the gravitational field caused by both, mass of the system and external mass. The simplest illuminating example (with the exact solution) is a point like body of mass $M_b$ in a constant external gravitational field $\boldsymbol{g}_{ext} = g_{ext}\boldsymbol{e}_z$, where $\boldsymbol{e}_z$ is the unit vector of the polar axis $z$ of spherical coordinates. Hence, in spherical coordinates $(r, \theta, \varphi)$ with the corresponding unit vectors $\boldsymbol{e}_r, \boldsymbol{e}_\theta, \boldsymbol{e}_\varphi$ we have $\boldsymbol{g}_N = -(GM_b/r^2)\boldsymbol{e}_r + g_{ext}cos\theta\boldsymbol{e}_r - g_{ext}sin\theta\boldsymbol{e}_\theta$.

## 2.2 Gravitational approximation of the quantum by ideal gas of non-interacting dipoles

Because of our very limited understanding of the quantum vacuum, the main question is how to find a reasonable approximation for the function $f_g(x_1, x_2, x_3)$. Unfortunately, absolutely the only option is to consider the quantum vacuum as a cosmological fluid composed of *non-interacting gravitational dipoles*, i.e. an *ideal gas* of virtual gravitational dipoles.

So, let us consider an ideal gas of virtual gravitational dipoles with the gravitational dipole moment $\boldsymbol{p}_g$ and energy $\varepsilon_g = -\boldsymbol{p}_g \cdot \boldsymbol{g}_N$ in a classical (Newtonian) gravitational field $\boldsymbol{g}_N$. Let us underscore that from a purely mathematical point of view, there is no difference between such a system of gravitational dipoles in an external gravitational field and well-known ideal paramagnetic gas, i.e. a system of non-interacting magnetic dipoles with the magnetic dipole moment $\boldsymbol{\mu}$ and energy $\varepsilon_\mu = -\boldsymbol{\mu} \cdot \boldsymbol{B}$ in an external magnetic field $\boldsymbol{B}$; the physical phenomena are different, but the mathematical equations are the same.

In principle an ideal gas of virtual gravitational dipoles (existing only as the quantum vacuum fluctuations) is different from an ideal gas of real (non-virtual) gravitational dipoles (which apparently do not exist).

If there were real gravitational dipoles, according to the statistical mechanics, the first step is to find the partition function, from which all macroscopic quantities (including the gravitational polarization density) can be obtained. In this case, the gravitational polarization density is a function of the well-known parameter $\beta \equiv 1/k_B T$ (where $k_B$ and $T$ denote respectively the Boltzmann constant and temperature), which is omnipresent in statistical mechanics.

However, in the case of virtual gravitational dipoles, instead of $\beta \equiv 1/k_B T$ a non-thermal parameter must be used, because, roughly speaking, quantum vacuum is a state of matter-energy at zero-temperature; to some extent like the quantum phase transitions. A quantum phase transition (see for instance Carollo 2018) occurs at zero temperature where thermal fluctuations are absent and instead the transition is driven by quantum fluctuations, which are tuned by variations in some non-thermal parameters; magnetic field is just one example of non-thermal parameters used instead of temperature $T$. Note that quantum phase transitions are transitions between two different states of ordinary matter immersed in the quantum vacuum; however we cannot exclude the existence of phase transitions between different states of quantum vacuum fluctuations. In principle, phase transition might exist as discontinuities of $P_g$ which, if they exist, are lost in the ideal gas approximation. However, at the present stage, the limited goal is only to show (using simplified studies) that the quantum vacuum as a source of gravity (through the mechanism of the gravitational polarization) deserves more profound consideration as a plausible candidate to explain phenomena usually attributed to dark matter and dark energy.

Let us consider the simplest case of a point-like body with mass $M_b$ immersed in the quantum vacuum, with additional simplification that energy $\varepsilon_g$ can have only two different values $\varepsilon_g = \pm p_g g_N$ (in other words the angle between $\boldsymbol{p}_g$ and $\boldsymbol{g}_N$ can have only two values (0 and $\pi$)). This is an illuminating case with exact solution. Because of spherical symmetry $\boldsymbol{g}_N/g_N = -\boldsymbol{r}_0$, where $\boldsymbol{r}_0$ denotes the unit vector of the radial coordinate $r$; consequently, the fundamental equation (1) for the effective gravitational charge density reduces to:

$$\rho_{qv}(M_b, r) = \frac{1}{r^2} \frac{d}{dr} \left[ r^2 P_g(M_b, r) \right] \qquad (4)$$

Just to remember, $P_g(M_b, r)$ denotes the magnitude of the gravitational polarization density $\boldsymbol{P}_g(M_b, r)$. Equation (4) leads to the following *effective gravitational charge* of the *gravitationally polarized quantum vacuum* within a sphere of radius $r$:

$$M_{gv}(M_b, r) = \int_0^r \rho_{qv}(M_b, r) dV = 4\pi r^2 P_g(M_b, r) \tag{5}$$

The corresponding acceleration caused by the quantum vacuum (which is now a source of gravity) is:

$$\boldsymbol{g}_{qv}(M_r, r) = -4\pi G P_g(M_b, r) \boldsymbol{r}_0 \tag{6}$$

Equation (6) shows that the magnitude of the gravitational acceleration caused by the quantum vacuum around a point like body is always smaller than acceleration $g_{qvmax} \equiv 4\pi G P_{gmax}$.

It is obvious that in the above equations $P_g(M_b, r) \leq P_{gmax}$ can be interpreted as the surface density of the effective gravitational charge of the quantum vacuum.

We can define a radius $R_{sat}$ so that

$$P_{gmax} = \frac{M_b}{4\pi R_{sat}^2} \Rightarrow R_{sat} = \sqrt{\frac{M_b}{4\pi P_{gmax}}} \tag{7}$$

The radius $R_{sat}$ is the radius at which the surface density corresponding to mass $M_b$ is equal to the maximum surface density $P_{gmax}$ that can be caused by the quantum vacuum around a point-like body. Roughly speaking, inside the sphere with radius $R_{sat}$ is region of saturation (in fact close to saturation). Just as an illustration, the radius $R_{sat}$ for a proton and a star as our sun is respectively a few times $10^{-14} m$ and about 10 thousand astronomical units AU, depending on the exact value of $P_{gmax}$ which is probably (Hajdukovic 2020) somewhere between $0.04$ and $0.06 \, kg/m^2$.

The corresponding solution for the magnitude of the gravitational polarization density is

$$P_g(M_b, r) = P_{gmax} \tanh\left(\frac{R_{sat}}{r}\right) \tag{8}$$

With the use of the explicit function $P_g(M_b, r)$ which is given by Eq.(8), equations (4), (5) and (6) become respectively, in principle testable, explicit functions for the effective gravitational charge density of the quantum vacuum $\rho_{qv}(M_b, r)$, the effective gravitational charge $M_{qv}(M_b, r)$ of the quantum vacuum within a sphere of radius $r$, and the gravitational acceleration $\boldsymbol{g}_{qv}(M_b, r)$ caused by quantum vacuum at distance $r$ from the body.

The ratio $R_{sat}/r$ in equation (8) has meaning only in the case of point-like body, but can be easily transformed to the following form that can be used in any gravitational field with the Newtonian magnitude $g_N$

$$\frac{R_{sat}}{r} = \frac{1}{r}\sqrt{\frac{M_b}{4\pi P_{gmax}}} \equiv \sqrt{\frac{g_N}{4\pi G P_{gmax}}} \equiv \sqrt{\frac{g_N}{g_{qvmax}}} \tag{9}$$

Consequently Eq.(8) which is valid only for a point-like body can be given more general form that is valid for any gravitational field $\boldsymbol{g}_N$.

$$\boldsymbol{P}_g = P_{gmax} \tanh\left(\sqrt{\frac{g_N}{4\pi G P_{gmax}}}\right) \frac{\boldsymbol{g}_N}{g_N} \tag{10}$$

According to equations (2) and (10) the unknown function $f_g(x_1, x_2, x_3)$ is equal to $tanh\left(\sqrt{g_N/4\pi GP_{gmax}}\right)$.

So far we considered the simplest case when energy $\varepsilon_g = -\boldsymbol{p}_g \cdot \boldsymbol{g}_N$ can have only two different values ($\varepsilon_g = \pm p_g g_N$). It is easy to generalize to the case when (as in the analogous case of paramagnetism) there are $2J + 1$ different energy levels (where J can be a positive half integer or integer number). In this more general case, hyperbolic tangent in equation (10) must be replaced by the corresponding Brillouin function. In the limit when $J$ tends to infinity, the Brillouin function $B_J(x)$ reduces to Langevin function $L(x)$ which describes a classical system with continuous spectrum of energy which can have any values in the interval $[-p_g g_N, p_g g_N]$. In general for a given $x$ and $J \geq 1$, $tanh(x) > B_J(x) > L(x)$; hence numerical values obtained using hyperbolic tangent and Langevin function can be respectively considered as an upper and lower bound.

If the angle between $\boldsymbol{p}_g$ and $\boldsymbol{g}_N$ can have 3 values $(0, \pi/2, \pi)$; note that this is analogous to 3 states of particle with spin one, instead of equation (10) the result is

$$\boldsymbol{P}_g = P_{gmax} \frac{2sinh\left(\sqrt{\frac{g_N}{4\pi GP_{gmax}}}\right)}{1 + 2cosh\left(\sqrt{\frac{g_N}{4\pi GP_{gmax}}}\right)} \frac{\boldsymbol{g}_N}{g_N} \equiv P_{gmax} B_1\left(\sqrt{\frac{g_N}{4\pi GP_{gmax}}}\right) \frac{\boldsymbol{g}_N}{g_N} \qquad (11)$$

Alternatively this can be written as

$$\boldsymbol{P}_g = P_{gmax} \frac{2cosh\left(\sqrt{\frac{g_N}{4\pi GP_{gmax}}}\right)}{1 + 2cosh\left(\sqrt{\frac{g_N}{4\pi GP_{gmax}}}\right)} tanh\left(\sqrt{\frac{g_N}{4\pi GP_{gmax}}}\right) \frac{\boldsymbol{g}_N}{g_N}$$

showing that the magnitude of the gravitational polarisation density corresponding to Eq. (10) is a little bit larger than the magnitude given by Eq. (11). As the difference is not very large, from the mathematical point of view (especially at this stage of toy models) it is better to use the simpler equation (10), while from the physical point of view the equation (11) is more plausible. Let us explain why.

Even in the ideal case when *all* quantum vacuum fluctuations are gravitational dipoles, many of them cannot contribute to the gravitational polarization of the quantum vacuum. For instance quark-antiquark pairs and electron-antielectron pairs would be also electric dipoles; consequently, such a weak force as the gravitational force cannot change their random orientation. In a way they would be sterile gravitational dipoles, gravitational dipoles that cannot be a source of gravity. Apparently, we must focus on fluctuations in which, *only particles without electric charge are included*; within the Standard Model of Particles and Fields the only basic particles without electric charge are neutrinos (with spin $s = \hbar/2$) and gauge bosons (photons, gluons and $Z^0$ boson, all with spin $s = \hbar$). If gauge bosons are somehow virtual gravitational dipoles, equation (11) seems more plausible; especially in the case of dipoles related to spin (described in subsection 3.3).

We are faced with a very intriguing situation. Quantum vacuum is an inherent part of quantum field theories, which are one of the most complex and deepest intellectual achievements of human mind. Opposite to the extreme complexity of quantum field theories is the extreme simplicity of the equation (10). It is absolutely astonishing and unbelievable that such a toy equation of a toy model can produce reasonable results. For instance this analytical solution is very close (Hajdukovic 2020) to ad hoc interpolating functions of MOND theory, which in a Galaxy well explains phenomena usually attributed to dark matter.

# 3. Different possibilities for the existence of gravitational dipoles

It is useful to remember that in electrodynamics an electric dipole is composed of a positive and a negative electric charge (for instance virtual electron-positron pairs in the quantum vacuum!); the essence is that there are positive and negative electric charges existing as electric monopoles. Contrary to this, a magnetic dipole is not composed of magnetic charges (magnetic monopoles) but is a result of the dynamics of electric charges. Hence, dipoles can be realised in fundamentally different ways and can exist with and without the existence of the corresponding monopoles.

Here, we present very different possibilities for the existence of gravitational dipoles; hopefully, other theoretical possibilities will be revealed in the near future. First, we point out what would be the best, i.e. an ideal kind of virtual gravitational dipole from the point of view of the gravitational polarization of the quantum vacuum; it is a dipole which interacts only through gravity. Next, because there are no candidates for ideal dipole in the Standard Model of Particles and Fields, we focus on possibilities within the Standard Model, suggesting gluons as good candidates. After that we consider possibilities for gravitational dipoles within the framework of General Relativity and virtual gravitational dipoles which can be directly attributed to the quantum vacuum regardless of General Relativity.

## 3.1 Sterile neutrino – a candidate for the best kind of virtual gravitational dipoles

From the point of view of the gravitational polarization of the quantum vacuum, it is obvious that *the best kind of gravitational dipole is a dipole which interacts only through gravity*. Other interactions can prevent the gravitational polarization; for instance, as already noticed at the end of Section 2, if a gravitational dipole is also an electric dipole, because of the fact that gravitational force is many orders of magnitude weaker than electromagnetic force, polarization is prevented.

Within the Standard Model of Particles and Fields there are no candidates for this best kind of gravitational dipole, simply because the Standard Model has no particles which participate *only* in gravitational interactions. However, there is strong theoretical and experimental motivation to complete the Standard Model of Particles and Fields with additional neutrinos, so called sterile neutrinos (for a recent review see Dasgupta&Kopp 2021), which are expected to interact only through gravity. It is likely that sterile neutrinos exist, and if it is confirmed by experiments, we will have a candidate for the best kind of virtual gravitational dipoles.

## 3.2 Gluon – the best candidate for virtual gravitational dipoles within the Standard Model?

In the absence of ideal candidates which interact only through gravity, the remaining good candidates within the current Standard Model of Particles and Fields, are only those constituents of the Standard Model which have no electric charge; hence, already known neutrinos, photons, gluons, $Z^0$ boson and the Higgs Boson. Among these candidates, as we suggest bellow, gluons might be the most promising ones. All other constituents of the Standard Model; quarks and antiquarks, charged leptons (electron, muon and tau) and their antiparticles, as well as $W^{\pm}$ bosons, have electric charge and cannot be aligned by gravity even if they are gravitational dipoles.

Strong interactions described by Quantum Chromodynamics and gravitational interactions described by General Relativity (which is not a quantum theory) look so different that any physical relation between them seems very unlikely; but who knows. As an encouragement to the open-minded attitude, let us point to a recently discovered correspondence (named *the double copy*) between scattering amplitudes (quantities related to the probability for particles to interact) in gravity and Quantum Chromodynamics (and in general gauge theories of the Standard Model). For instance, the double copy expresses graviton tree amplitudes (i.e. a tree-level Feynman diagrams) in terms of sums of products of gluon tree amplitudes. No one knows if these mysterious connections between gauge and gravity theories (in particular Quantum Chromodynamycs and gravity) have a physical meaning, or it is just a formal, mathematical similarity between otherwise completely unrelated physical systems. In any case (see for instance two recent reviews White 2018, Bern et al. 2022) it is intriguing that, within the double-copy framework for gravity, gravity directly follows from Quantum Chromodynamics, as stated already in the amusing title The double copy: gravity from gluons" (White 2018).

Let us note that we still do not know if by its nature gravity (as other interactions) should be described by a quantum theory, but if yes, the quantum vacuum must contain fluctuations related to gravity.

In brief, there are a lot of unknowns and hence a lot of room for virtual gravitational dipoles within the quantum vacuum of the current Standard Model. In addition there are friendly extensions of the Standard Model which do not challenge its essence (for instance, addition of sterile neutrinos and the corresponding quantum vacuum fluctuations) and possibility that quantum vacuum contains fluctuations related to gravity if gravity is a quantum theory. We do not advocate any of these possibilities; the point is that the quantum vacuum as a source of gravity through the gravitational polarisation of the quantum vacuum is a so precious scenario that, despite respectable theoretical mainstream arguments against gravitational dipoles, we must stay open minded even (or especially) for the unlikely possibilities.

### 3.3 Gravitational dipoles within the framework of General Relativity

There are exciting and to some extent surprising possibilities, that gravitational dipoles (including virtual gravitational dipoles in the quantum vacuum) might exist within the framework of General Relativity.

### 3.3.1 Gravitational dipole moment in General Relativity with torsion

The Weak Equivalence Principle (telling us that inertial mass and gravitational charge is the same thing, what is often expressed as the universality of the free fall) is the cornerstone of General Relativity. Consequently, negative gravitational charge and gravitational dipoles analogous to electric dipoles cannot exist within the framework of General Relativity.

However, the existence of microscopic gravitational dipoles which are not similar to electric dipoles (for instance, to some extent they can be similar to magnetic dipoles existing without the existence of magnetic charges) remains possible. In fact, we already have a theoretical example of such a gravitational dipole in a simple extension of General Relativity; an exact solution (Hammond 2020) to the gravitational field equations with torsion shows that there is a dipole gravitational field, even though there is no negative gravitational charge. Let us underscore that the corresponding gravitational dipole moment $\mathbf{p}_g$ is related to the intrinsic spin $\mathbf{S}$ (which is a fundamentally quantum phenomenon); from the original paper (Hammond 2020) it is easy to get the following relation for the magnitude of $\mathbf{p}_g$:

$$p_g(S) = \frac{12}{A} \frac{GS^2}{c^4} \qquad (12)$$

In the above equation $A$ is a constant surface which is not determined in the original paper; hence there is an unknown constant of proportionality $1/A$ (which in original paper was denoted by λ).

While no one noticed it, the Eq.(2) suggests that all virtual particles and antiparticles in the quantum vacuum (hence all quarks and leptons with spin $S = \hbar/2$, and gauge bosons with spin $S = \hbar$) are virtual gravitational dipoles, while Higgs Boson which is scalar boson with spin $S = 0$ is not a gravitational dipole. Hence, quantum vacuum fluctuations might be gravitational dipoles not because of the gravitational charges of the opposite sign, but because of spin and torsion.

In our previous work it was shown that (if it exists) the magnitude of the gravitational dipole moment of a quantum vacuum fluctuation (composed of two gravitational charges of the opposite sign) cannot be larger than $\hbar/c$ and as a satisfactory value we have used:

$$p_g = \frac{1}{2\pi}\frac{\hbar}{c} \tag{13}$$

While dipoles given by Eq.(12) and Eq.(13) have very different physical nature, if we assume that they have the same magnitude, the gravitational field caused by one dipole would be indistinguishable from the gravitational field caused by the other dipole. Of course, we can always select the unknown constant A in Eq.(12) so that magnitude of two dipoles is the same (or has a very close value). However, we are not interested in such a mathematical choice of the constant $A$, without any physical significance.

However, we have noticed that there is a surprising coincidence, which hopefully has physical significance. If we assume that constant $A$ is the surface of a "Planck sphere", i.e. $A = 4\pi l_p^2$ (where $l_p$ is the Planck length), Eq.(12) reduces respectively to:

$$p_g\left(S = \frac{\hbar}{2}\right) = \frac{3}{4\pi}\frac{\hbar}{c}; \quad p_g(S = \hbar) = \frac{3}{\pi}\frac{\hbar}{c} \tag{14}$$

It is astonishing and intriguing that dipoles related to intrinsic spin and dipoles related to the hypothetical existence of positive and negative gravitational charge in the quantum vacuum fluctuations have nearly the same magnitude.

In order to get the gravitational polarization of the quantum vacuum it is sufficient that only one of these two possibilities is realized in nature; in particular, in spite of different physical nature of gravitational dipoles, results obtained in the previous papers (Hajdukovic 2020, Hajdukovic and Walter 2021) remain valid. And, as we will show, there are other scenarios for the existence of gravitational dipoles.

It is useful to underscore the following two things.

First, there is no fundamental reason, apart from simplicity, to assume (as it is assumed in General Relativity) that space-time is torsionless. The main feature of General Relativity is that mass-energy generates the curvature (and interacts with the curvature) of space-time, but the torsion of space-time is zero. Torsion is the most natural extension of General Relativity and apparently the best way to include effects related to intrinsic spin of particles (i.e. the best way of incorporating spin in a geometric description); it is quite possible that intrinsic spin generates the torsion (and interacts with the torsion) of space-time in an analogous way as mass-energy generates the curvature. Without guidance from experimental measurements, there are a number of open theoretical possibilities; we need the experimental tests at the intersection of quantum physics and General Relativity and hopefully we will have such tests in a visible future as for instance a recently proposed measurement (Fadeev et al. 2022) of general relativistic precession of intrinsic spin using a ferromagnetic gyroscope with unprecedented sensitivity.

Second. The general opinion is that in General Relativity the impact of curvature is many orders of magnitude larger than the impact of torsion. For sure this is true if we assume a gravitationally featureless classical vacuum, but it is questionable (and perhaps wrong) in the quantum vacuum in which, because of spin, each virtual particle might be a virtual gravitational dipole with a significant impact of the gravitationally polarized quantum vacuum. Hence, if the polarized quantum vacuum is taken into account, the impact of torsion might be even larger than the impact of curvature.

### 3.3.2 Gravitomagnetic dipoles

Gravitomagnetism (this name suggests the existing analogy with electromagnetism) is an approximation of General Relativity at low velocities and weak gravitational fields. The key point is that gravitomagnetic equations (which follow from General Relativity) are analogous to electromagnetic Maxwell equations.

The main difference between Newtonian theory of gravity and Gravitomagnetism (which can be considered as a general relativistic extension of gravity) is that, in addition to the ordinary Newtonian gravitational field (described by Newtonian acceleration $\boldsymbol{g}_N$), there is a second, gravitomagnetic field $\boldsymbol{B}_{gm}$ caused by moving masses (hence, $\boldsymbol{B}_{gm}$ is the gravitational analogue of magnetic field $\boldsymbol{B}$ produced by the motion of electric-charges).

If we trust General Relativity we should trust the gravitomagnetism as well. However it is no longer a question of trusting. The Gravity B probe experiment (Everitt 2011) has confirmed the existence of both, the gravitomagnetic field caused by the Earth's rotation, and the gravitomagnetic field caused by the Earth's orbiting around Sun.

From the point of view of the paradigm that quantum vacuum is a source of gravity (through the mechanism of the gravitational polarization) the most important result is that all rotating bodies (planets, stars…) produce a gravimagnetic field related to the angular momentum of the rotating body. Hence a rotating body is a gravitomagnetic dipole (analogous to magnetic dipole). Knowing this, it seems plausible that all particles with intrinsic spin are gravitomagnetic dipoles. Consequently, the quantum vacuum might be full of virtual gravitomagnetic dipoles which can be aligned (i.e. quantum vacuum can be polarised) by the gravitomagnetic field of bodies immersed in the quantum vacuum and moving through it.

3.3.2.1 Gravitomagnetism and virtual gravimagnetic dipoles instead of dark matter?

Let us underscore that there are so far unsatisfactory attempts (see Ciotti 2022 and references therein) to explain by gravitomagnetism the galactic phenomena usually attributed to hypothetical dark matter. An important question that deserves attention of authors of these studies is if the quantum vacuum with virtual gravitomagnetic dipoles and the corresponding gravitomagnetic polarisation of the quantum vacuum can lead to a satisfactory description of galactic dynamics by gravitomagnetism, without need for dark matter.

## 3.4 Gravitational dipoles directly attributed to the quantum vacuum fluctuations

Regardless of General Relativity, virtual gravitational dipoles can be directly attributed to the quantum vacuum fluctuations.

### 3.4.1 Gravitational dipoles composed of gravitational charges of the opposite sign

We know that there are quantum vacuum fluctuations which are virtual electric dipoles (for instance fluctuation composed of a quark and its antiquark, or fluctuation composed from an electron and antielectron). However, there are also fluctuations (for instance a neutrino-antineutrino pair or photons) which are not electric dipoles. In brief, we can say that virtual electric dipoles are a subset of all quantum vacuum fluctuations.

The same may be true for virtual gravitational dipoles (if they are eventually composed of gravitational charges of the opposite sign). In principle there are 3 different possibilities:
- *All* quantum vacuum fluctuations are virtual gravitational dipoles
- There is only a subset of quantum vacuum fluctuations which are virtual gravitational dipoles
- There are no quantum vacuum fluctuations which are gravitational dipoles

In previously published papers, the working hypothesis that *all* quantum vacuum fluctuations are virtual gravitational dipoles was accompanied with an excellent working model of a virtual gravitational dipole, *analogues to electric dipole*, i.e. composed of two gravitational charges of the opposite sign; with positive and negative gravitational charge respectively assigned to virtual particles and antiparticles.

Regardless if it is wrong or correct, this model of virtual gravitational dipole is the best one from the point of view of a pedagogical presentation. Second, even if wrong, this model can be used because the consequences of the gravitational polarization of the quantum vacuum, mainly do not depend on the physical nature of dipoles but only on their existence, magnitude and number density (which is in principle determined by the density of the corresponding quantum vacuum fluctuations); for instance (as we noticed above) a dipole related to spin and torsion and a dipole related to the opposite gravitational charge, can produce similar consequences. In addition (while it is not a scientific reason) it was a lovely and beautiful hypothesis; physics is already poetic, but it would be much more poetic if there is gravitational repulsion between matter and antimatter.

Unfortunately, this otherwise very useful model of gravitational dipole has created a very strong prejudice that the gravitational polarization of the quantum vacuum is possible *only if* there is a universal gravitational repulsion between matter and antimatter. No, things are much more complex; just one example of complexity is that, as we have seen, gravitational dipoles might exist within the framework of General Relativity without any reference to antimatter. However, as we will argue bellow, even gravitational dipoles related to antimatter cannot be excluded on the basis of the fact that atoms and antiatoms have gravitational charge of the same sign (as revealed by antimatter gravity experiment at CERN, Anderson *et al.* 2023).

Before we continue let us also point to the old but unnoticed warnings which were aimed to prevent prejudices. The following is a quotation from Hajdukovic 2016 and Hajdukovic 2020: „Apparently, the simplest and the most elegant realization of this hypothesis is if particles and antiparticles have gravitational charge of the opposite sign; *of course, nature may surprise us with a different realization of the gravitational dipole-like behaviour of the quantum vacuum*". In addition the "philosophy" of the approach was clearly stated: „*it is more important and productive to reveal the physical consequences of the hypothesis than to enter into the exchange of purely theoretical arguments*".

After these clarifications, let us focus on the ALPHA-g experiment at CERN.

### 3.4.2 Antimatter gravity experiment at CERN

Recently, incredibly sophisticated, difficult and fascinating experiments of the ALPHA collaboration, for the first time in human history, have demonstrated that atoms and antiatoms (more precisely hydrogen and antihydrogen) fall in the same direction, while a difference in magnitude of acceleration remains possible and would be subject of further studies. It is very likely that this seminal result will be confirmed by other experiments with antihydrogen at CERN (AEGIS and GBAR).

In brief, the ALPHA-g experiment has revealed that both *atoms* and *antiatoms* fall towards the Earth, or equivalently that *atoms and antiatoms have gravitational charge of the same sign*. We face the immediate question what are the consequences of this empirical evidence, for the existence of virtual gravitational dipoles composed of two gravitational charges of the opposite sign.

The answer to the above question depends on the extent to which *a virtual particle-antiparticle pair is distinct from its "real" analogue*. The key point is that physical proprieties of a quantum vacuum fluctuation and the analogous "real" particle are not identical; the most obvious is a different lifetime, but there is difference in mass as well. For instance, a real neutral pion $\pi^0$, i.e. the lightest quark-antiquark meson, which is made up of a *superposition* (an up, anti-up pair and a down, anti-down pair) has to some extent different physical properties from the corresponding quantum vacuum fluctuation which is an up, anti-up pair or a down, anti-down pair. The same must be valid in all other cases; one more example is that a gluonium (a composite system made of two gluons) is not identical to its virtual counterpart in the quantum vacuum.

From the point of view of the Standard Model (i.e. from the point of view of the quantum theory) the differences are not radical; the most important, particles and their counterparts in the quantum vacuum have the same spin and charges of all three interactions (strong, electromagnetic and weak interactions). However, from the point of view of gravity, because of the cosmological constant problem, the open possibility is that differences are radical and crucial. Hence, because of the cosmological constant problem there is no certainty if we can extend to "virtual" matter in the quantum vacuum, the gravitational results of the ALPHA-g experiment which were obtained for real matter and antimatter.

If we neglect the cosmological constant problem and extend ALPHA-g results to the quantum vacuum, we can only say that matter and antimatter have the gravitational charge of the same sign within *the quark sector of the Standard Model* of Particles and Fields. In principle it remains to test *antimatter gravity* in the other two sectors of the Standard Model, *leptonic* sector and gauge bosons sector. In near future, we will have tests of antimatter gravity in leptonic sector, with positronium and muonium, while, at this stage of scientific knowledge it looks unimaginable to study gravity in the sector of gauge bosons like gluons and $Z^0$.

Hence, when the quantum vacuum is in question, from the purely experimental point of view (theoretical opinions are different topics), ALPHA-g disproves only the existence of virtual gravitational dipoles in the quark sector, i.e. for the quantum vacuum fluctuations that include only quarks and antiquarks. This disproval of one particular kind of virtual gravitational dipoles is valid only if for some reasons we can neglect the cosmological constant problem. By the way, let us remember that quarks have electric charge; hence even if these gravitational dipoles are not excluded, they are not significant for the gravitational polarization.

The only, but perhaps not crucial shortcoming of the ALPHA-g experiment is that behaviour of antihydrogen in the gravitational field of the Earth was not directly compared with the behaviour of hydrogen. In principle the experiment must be done with both hydrogen and antihydrogen so that results can be compared. However, the experiment with hydrogen is not possible at the ALPHA-g experiment, because sophisticated facilities do not permit replacing the source of antihydrogen with a source of hydrogen. Hence, instead of a direct comparison with hydrogen, experimental results for antihydrogen were compared with *numerical simulations of trajectories of hydrogen under ALPHA-g conditions*; however, the real conditions in the vertical tube might be different from our estimation of the sensitive mixture of electromagnetic and gravitational field within the tube.

## 3.5 Virtual gravitational dipoles as a subset of the quantum vacuum fluctuations

According to the above discussion the ALPHA-g experiment questions only the most general statement that *ALL* quantum vacuum fluctuations are virtual gravitational dipoles, but doesn't exclude the possibility that virtual gravitational dipoles are only a subset of the quantum vacuum fluctuations.

However there is an amusing and fundamental fact that in principle, experiments with all kinds of matter-energy in the Universe, cannot completely exclude the existence of virtual gravitational dipoles in the quantum vacuum. Imagine that we have complete experimental evidence that there is only positive gravitational charge in the Universe, and that all results can be extended to the corresponding counterparts in the quantum vacuum. Even in such a case, there are quantum vacuum fluctuations which might be virtual gravitational dipoles. The reason is that already within the current Standard Model of Particles and Fields (hence without introducing something new) *there is possibility to have quantum vacuum fluctuations without real counterparts*.

In simple words, according to Quantum Chromodynamics (our best theory of strong interactions) *all constituents of matter-energy in the Universe are colourless*. Hence, all our experiments are limited to colourless matter-energy, while in the quantum vacuum there are fluctuations which are not colourless. It is obvious that in a Universe of colourless matter-energy, there are no counterparts for the quantum vacuum fluctuations which have colour.

Let us make this clearer to people not familiar with Quantum Chromodynamics.

In Quantum Electrodynamics there are only two kinds of electric charge that we call positive and negative electric charge; electric charges interact through the exchange of photons (in other words the photon mediates the electrodynamics force, or the photon is the force carrier for electromagnetism). In the case of Quantum Chromodynamics, i.e., in the case of strong interactions, distinct from electric charge there is *strong charge* (called *colour charge*); strong charges interact through the exchange of gluons. Hence, in Quantum Chromodynamics gluons play a role analogous to photons. You may think about Quantum Chromodynamics as a generalisation of Quantum Electrodynamics with 6 strong charges; three colour charges (red, green and blue for quarks) and three anti-colour charges (anti-red, anti-green and anti-blue for antiquarks) while the 8 existing gluons are *roughly speaking* bicolour particles with a colour and anti-colour charge.

By definition a *colourless* state is a state with a net colour charge equal to zero. Because of a property of the strong interaction called *colour confinement*, free particles must have a colour charge of zero, i.e. *"real" particles are always colourless* (strictly speaking this is true only below a very high temperature above which quark-gluon plasma can exist). Protons, neutrons, electrons, photons, mesons, everything is colourless.

A colour and its anti-colour (red and anti-red, green and anti-green, blue and anti-blue) is colourless, while for instance red and anti-green is not colourless. Mesons which are always composed of two quarks must have the opposite colour of quarks in order to be colourless. Mesons with a red and anti-red strong charge exist, while mesons with a red and anti-green strong charge do not exist as real particles, but might exist as quantum vacuum fluctuations. For particles as protons, neutrons and their antiparticles, which are composed of three (valence) quarks, colourless is only the state with all 3 quarks having different colours (or anti-colours).

Let us underscore that all eight gluons corresponding to $SU(3)$ symmetry in Quantum Chromodynamics have non-zero colour charge; hence *there are no colourless gluons*.

All this gives an idea of complexity and the value of the pragmatic "philosophy" that at this stage, it is more important and productive to reveal the physical consequences of the hypothesis than to enter into the exchange of the endless purely theoretical arguments.

## 4. Three new potential phenomena considered within the framework of the gravitational polarization

If confirmed, three potential astronomical phenomena that emerged this year will become an unprecedented challenge for the Standard ΛCDM Cosmology. These three potential phenomena include:
- A surprisingly fast formation of the first stars and galaxies (Ferreira et al.2023; Labbé et al. 2023).
- The existence of massive relic galaxies, as NGC 1277 galaxy, which is dark matter deficient (Comerón et al. 2023).
- Non-Newton gravity in the internal dynamics of Wide Binary Stars (Chae 2023, Hernandez 2023).

In this section we present the first indications that all these phenomena can be explained within the framework of the gravitational polarization of the quantum vacuum.

In the following calculations we will use numerical values that were used in the previous publications: $g_{qvmax} \equiv 4\pi G P_{gmax} = 5 \times 10^{-11}\, m/s^2$ and $4\pi P_{gmax} \approx 0.75\, kg/m^2$; however, because the exact values are not known, for comparison we will also calculate results for slightly different values. As an illuminating example of plausibility of these values, let us calculate dark matter mass (i.e. the effective gravitational charge density of the quantum vacuum) of our galaxy within the sphere of 260kpc, and local dark matter density (i.e. density at the distance of about 8kpc from the centre of our galaxy). From equations (5), (7) and (8), dark matter mass within radius $r$ is

$$M_{dm}(M_b, r) = 4\pi P_{gmax} r^2 \tanh\left(\frac{1}{r}\sqrt{\frac{M_b}{4\pi P_{gmax}}}\right) \quad (15)$$

For large $r$ the above equation reduces to

$$M_{dm}(M_b, r) \approx \left(\sqrt{4\pi P_{gmax} M_b}\right) r \quad (16)$$

For $r = 260\, kpc$, and astronomical estimate of baryonic mass mass $M_b \approx 0.6 \times 10^{10} M_\odot$,, the result is

$$M_{dm}(M_b = 1.2 \times 10^{41} kg, r = 260 kpc) = 1.2 \times 10^{12} M_\odot = 2.4 \times 10^{42} kg \quad (17)$$

A surprisingly good prediction from a double-toy model; a toy model considering the quantum vacuum as an ideal gas of virtual gravitational dipoles, and a toy model approximating the baryonic mass in galaxy with a point-like body.

Concerning dark matter density at distance of about $8 kpc$ from the centre of our galaxy, let's give an easy, back of the envelope calculation. Distance of $8 kpc$ is smaller than the saturation radius $R_{sat} = \sqrt{M_b/4\pi P_{gmax}} \approx 13 kpc$; hence, according to Eq. (8), $P_g(M_b, r)$ can be approximated by $P_{gmax}$ and Eq.(4) simply reduces to

$$\rho_{qv}(M_b, r) = \frac{2 P_{gmax}}{r} \Rightarrow \rho_{qv}(r = 8 kpc) \approx 7 \times 10^{-3} \frac{M_\odot}{pc^3} \quad (18)$$

This can be compared for instance with the value $8.8 \times 10^{-3} \, M_\odot/pc^3$ from one of sophisticated astronomical empirical studies ().

### 4.1 Non-Newton gravity in the internal dynamics of Wide Binary Stars

Recently, internal kinematics of wide binary populations, carefully selected from the Gaia DR3 database, was studied statistically (Chae 2023, Hernandez 2023). The conclusion is that Newtonian gravitational force between stars is only a fraction of the total gravitational force. The additional, seemingly non-Newtonian force of gravity cannot be attributed to dark matter as in galaxies; hence authors concluded that the additional force can only be explained by modification of gravity within MOND paradigm. However, according to the paradigm of the gravitational polarization of the quantum vacuum, the additional gravitational force can be completely explained by the Newton law, including the quantum vacuum as a so far forgotten Newtonian source of gravity.

A system of two stars immersed in the quantum vacuum at mutual distance $D$ is described by Equations (A1), (A2) and (A3) given in the Appendix; of course in addition to $D$ it is necessary to know masses of stars (denoted $M$ and $m$). However, because of the inevitable statistical nature of the astronomical studies, there is no possibility to compare the theoretical predictions with the observed dynamics of individual systems. So, it is necessary to extend the statistical procedure to include the theory of the gravitational polarisation. In principle it is not difficult but demands huge amount of numerical calculations based on already existing simulations for about 26000 wide binaries in Chae 2023, and similarly in Hernandez 2023. Here we only give the first indication that quantum vacuum can explain the observations.

Chae 2023, considered a subset of wide binary stars that satisfy the following condition

$$G_N^* = \frac{GM_{tot}}{D^2} \approx 10^{-10.15} \, m/s^2 \approx 7.1 \times 10^{-11} \, m/s^2 \quad (19)$$

Here, $M_{tot} = M + m$, denotes the total mass of the wide binary, while instead of the original notation $G_N$ we use $G_N^*$ in order to distinguish it from notation used in our paper for the magnitude of gravitational acceleration caused by both stars. Note that binaries with very different total mass and distance can satisfy this condition.

For this subset of wide binaries, the observed ($g_{obs}$) to Newton-predicted ($g_{pred}$) acceleration ratio is

$$\frac{g_{obs}}{g_{pred}} = 1.43 \pm 0.6 \quad (20)$$

In order to avoid tedious numerical calculations, let us consider the case $m \ll M$; consequently $M_{tot} \approx M$ and $g_N \approx g_N^*$. Hence, according to equations (6), (10) and (11)

$$\frac{g_{obs}}{g_{pred}} \equiv \frac{g_{obs}}{g_N} \approx 1 + \frac{g_{qvmax}}{g_N} \begin{cases} tanh\left(\sqrt{\frac{g_N}{g_{qvmax}}}\right) \\ \\ \dfrac{2sinh\left(\sqrt{\frac{g_N}{g_{qvmax}}}\right)}{1 + 2cosh\left(\sqrt{\frac{g_N}{g_{qvmax}}}\right)} \end{cases} \quad (21)$$

The numerical results corresponding to two different functions in Eq.(21), for slightly different values $g_{qvmax}$ are given in Table 1. Because of possible relation with spin $S$, results obtained using hyperbolic tangent are given in column denoted $S = 1/2$, while column $S = 1$ corresponds to the other function.

Table 1. The ratio $g_{obs}/g_{pred}$ for slightly different values $g_{qvmax}$.
Column $S$=1/2 corresponds to the hyperbolic tangent in Eq. (21), while column $S$=1 corresponds to the other function.

|  | $S = 1/2$ | $S = 1$ |
|---|---|---|
| $g_{qvmax} = 5 \times 10^{-11} \, m/s^2$ | 1.59 | 1.46 |
| $g_{qvmax} = 4.5 \times 10^{-11} \, m/s^2$ | 1.54 | 1.426 |
| $g_{qvmax} = 4 \times 10^{-11} \, m/s^2$ | 1.49 | 1.39 |

Note that column $S = 1$, corresponds much better to the empirical evidence given by Eq. (20); in fact empirical evidence, i.e. ratio 1.43 in Eq.(20) is identical to our ratio 1.426 that is marked in Table 1. This should be considered as motivation for a full statistical and numerical study based on the exact solution for two point-like bodies given in Appendix A.

### 4.2 Dark matter deficient NGC 1277 galaxy

According to a recent paper (Comerón et al. 2023), the massive relic galaxy NGC 1277 is dark matter deficient. Within the framework of the $\Lambda$ cold dark matter ($\Lambda CDM$) cosmology, present-day galaxies with stellar masses $M_* > 10^{11} M_\odot$ should contain a sizable fraction of dark matter within their stellar body, but it seems that the content of dark matter within $5R_e$ (5 effective radii) of NGC 1227, is one order of magnitude smaller than predicted by dark matter models. More precisely, dark matter fraction within 5 Re was found to be $f_{DM}(5R_e) < 0.05$ while the expected fraction is about 10 times larger. In other words, there is only a small deviation from Newtonian gravity and only a small amount of dark matter must be added in order to fit observations.

Let us consider NGC 1227 galaxy within the framework of the gravitational polarization of the quantum vacuum. We already know that the amount of the effective gravitational charge in a volume (in this case the volume of a sphere with radius equal to $5R_e$) depends on both, the immersed baryonic mass and the way in which baryonic mass is distributed; different distributions of the same baryonic mass $M_*$ can produce very different amounts of the effective gravitational charge. Hence, from the point of view of the gravitational polarisation, there is something special in the distribution of baryonic matter in NGC 1227 galaxy; a tiny amount of the effective gravitational charge density is a consequence of particularity of distribution of baryonic matter.

What is unusual in distribution of baryonic mass in NGC 1227? The total stellar mas of the galaxy is $M_* \approx 1.8 \times 10^{41} M_\odot$ with a supermassive black hole (with a mass of a few billions solar masses) in the centre of the galaxy.

The key point is that NGC 1227 is an extremely compact galaxy. The effective radius (or half-light radius) of the galaxy is $R_e \approx 1.2 kpc \approx 3.7 \times 10^{19} m$, and consequently $5R_e \approx 6 kpc \approx 1.85 \times 10^{20} m$; this effective radius is a few times smaller than for other galaxies with similar mass. The stellar mass within is $X_*(5R_e) \approx 0.84 M_* \approx 3 \times 10^{41} M_\odot$; it is as if all baryonic mass of Milky Way is within 6kpc from the centre, deep inside the solar orbit. Because of such a compactness and very large mass of the central black hole, the saturation radius (see equation 7) of the black hole is about $10^{20} m$; saturation radius of the supermassive black hole is larger than $2R_e$. The black hole only can assure a region of saturation within a sphere of radius $2R_e$ and apparently it is reasonable to approximate the gravitational polarisation of the quantum vacuum by a point-like body.

Now, according to equations (5) and (8), using the value $g_{qvmax} = 4.5 \times 10^{-11} m/s^2$ (i.e. the same value which is apparently the best choice in the case of wide binary stars) we have

$$f_{DM}(5R_e) \approx \frac{\frac{4\pi P_{gmax}(5R_e)^2}{X_*(5R_e)} tanh\left[\frac{R_{sat}(X_*(5R_e))}{5R_e}\right]}{1+\frac{4\pi P_{gmax}(5R_e)^2}{X_*(5R_e)} tanh\left[\frac{R_{sat}(X_*(5R_e))}{5R_e}\right]} \approx 0.07 \qquad (14)$$

Note that in the above expression hyperbolic tangent is with high accuracy equal to 1 ($\approx 0.9985$) because the saturation radius corresponding to baryonic mass $X_*(5R_e)$ is $\approx 7 \times 10^{20}$m, i.e. about 3.6 times larger than $5R_e$. There is no a significant difference if instead of hyperbolic tangent, we use the function given by Eq.(11), i.e. the Brullouin function for $J = 1$ (instead of $tanh(3.6) = 0.998$, $B_1(3.6) = 0.972$).

### 4.3 A surprisingly fast formation of the first stars and galaxies

Preliminary results of the James Webb Space Telescope suggest that formation of structures in the Universe was much faster (Ferreira et al.2023; Labbé et al. 2023) than it should be according to the standard $\Lambda CDM$ cosmology. We present only a qualitative description of how the gravitational polarization of the quantum vacuum may contribute to a faster creation of stars and galaxies; a quantitative description demands sophisticated numerical studies.

We know that before the beginning of the structure formation in the Universe, the content of the Universe was a rarefied, nearly uniformly distributed gas composed of hydrogen and helium. However, distribution of gas was not perfectly uniform; there were cosmic "clouds" with a small but crucial overdensity. Structure formation in the Universe is a result of gravitational amplification of these initial density perturbations (initial clouds of cosmic gas).

Let us note that in the initial gas (roughly at redshift $z \approx 1000$), the mean distance between two nuclei which are the first neighbours was about $10^{-3}m$, i.e. about 10 orders of magnitude larger than the saturation radius of a nucleus ($R_{sat} = \sqrt{m/4\pi P_{gmax}}$) defined by Eq.(7)

In principle a single nucleus of hydrogen or helium can create its own halo of the polarised quantum vacuum (Hajdukovic 2020); this is possible if there is a region around nucleus (a region whith a size much larger than $R_{sat}$) in which the gravitational field of the nucleus is much stronger than the external gravitational field caused by the rest of the Universe. Because of the antiscreening effect of the polarized quantum vacuum, in a way described by Eq.(5), such a nucleus can have the effective gravitational charge perhaps a few orders of magnitude larger than its own gravitational charge.

The best conditions for the existence of halos of the polarized quantum vacuum corresponding to individual atoms are in a Universe which is entirely an ideal, rarefied gas with constant density. Hence, within a nearly homogenous initial cosmic gas, atoms should have their own halos of the polarized quantum vacuum; consequently the effective gravitational charge of atoms is much larger and gravitational force between them much stronger. This may help a much faster increase of compactness of an initial cloud; i.e. clouds would be faster ready for creation of stars.

Of course, if a single point-like body like a star is created from an initial gas of atoms having individual halos of the polarized quantum vacuum, during that process individual halos must be lost; a transition must exist from individual halos of atoms to a "collective" halo, i.e. the halo of the polarized quantum vacuum around the new formed star. In principle, a Universe friendly for the existence of individual atomic halos would become unfriendly for such existence (for instance, an atom cannot create a halo in competition with a much stronger gravitational field of Sun or Earth) ;

however, the main initial job, more compact cosmic clouds will probably be finished. In such a way, stars and massive galaxies may be eventually formed faster than in the Standard $\Lambda CDM$ cosmology.

## 5. Complexity of the gravitational polarization of the quantum vacuum

The concept of the gravitational polarisation of the quantum, caused by the well-known Standard Model matter which is immersed in it, is very simple and elegant but the result of polarisation is very complex and analytical solutions are possible only for the simplest systems, like a single point-like body, two point-like bodies and some particular distributions of matter.

A single point-like body is surrounded by a spherically symmetric *halo* of the polarised quantum vacuum; this *halo* of the polarized quantum vacuum (apparently) has potential to explain phenomena usually (and maybe wrongly) attributed to dark matter or modification of gravity.

Two point-like bodies (Hajdukovic and Walter 2021) are surrounded by a cylindrically symmetric distribution of the effective gravitational charge density of the polarised quantum vacuum. The main feature is that in the case of two bodies, the effective gravitational charge density can be negative (for more details see the Appendix A). Hence, in the "ocean" of the prevailing positive effective gravitational charge density in the Universe, there are many islands of negative effective charge density; in particular (of course if the gravitational polarization exists) it is valid for our Solar system as well.

Now, as an extreme but illuminating example of mass distribution, let's imagine an infinite plane of mass, with a constant surface mass density $\sigma$ (hence $\sigma$ is mass per unit area). Within the framework of the Newtonian gravity (using the Gauss's law of gravity), it is easy to show that the gravitational acceleration $\boldsymbol{g}_N$ caused by the plane, is directed towards the plane and has a constant magnitude $g_N = 2\pi G\sigma$ independent from distance $z$ from the plane. Hence, $g_N$ is a constant vector (with opposite direction on different sides of the plane).

If you see this result for the first time you may be surprised; the acceleration is independent of the distance from the plane! While this result is a consequence of the Newton's inverse square law, it is not "visible"; it is masked by a particular distribution of matter. This extreme example shows how important is the way in which a given mass is distributed.

Of course, the constant gravitational field $\boldsymbol{g}_N$ produces a constant gravitational polarization of the quantum vacuum, i.e., the vector $\boldsymbol{P}_g$ of the gravitational polarization density is constant as well. Consequently, according to the fundamental equation $\rho_{qv} = -\boldsymbol{\nabla} \cdot \boldsymbol{P}_g$, density $\rho_{qv}$ as divergence of a constant vector $P_g$ must be zero everywhere. Despite the existence of the gravitational polarization (it can even be saturation at all points), quantum vacuum *outside* the plane is not a source of gravity.

Between examples of a point-like body and infinite plane of matter, there is one more illuminating example, an infinitely long line of mass with constant mas per unit length, producing a different pattern of the gravitational polarization of the quantum vacuum.

The conclusion is that the effects of the gravitational polarization of the quantum vacuum significantly depend on the way in which the Standard Model matter is distributed within the quantum vacuum.

Let us remember that astronomical observations suggest that distribution of matter in the present-day Universe has a web-like configuration called the Cosmic Web. Roughly speaking the Cosmic Web has the following components: nodes (or knots), filaments, sheets (or walls) and voids. Gravitational polarization caused by these very different components (galactic filaments are very elongated quasi one-dimensional structures, sheets are quasi planar structures, nodes are in fact cluster regions) must be very different.

Because of many unknowns and such a universal complexity of the gravitational polarisation it is absolutely premature to claim (Banik&Kroupa 2020) that the gravitational polarization within the Solar System is dismissed by empirical evidence. For instance, authors neglected at that time unknown existence of regions with a negative effective gravitational charge. By the way, authors recognised that theory which they considered to be dismissed within the Solar System apparently works well at galactic scale.

## 6. Outlook

After nearly one century we do not know the answer to the most fundamental question if the quantum vacuum is a source of gravity or not.

It is likely that the quantum vacuum is a major source of gravity in the Universe and this is probably a time bomb on which sit both the Standard $\Lambda CDM$ cosmology and its current alternatives that attempt modification of gravity.

The essence of the cosmological constant problem is that from our best theories (The Standard Model of Particles and Fields and General Relativity) comes an unbelievable wrong theoretical prediction: *the quantum vacuum is at least a 40-orders-of-magnitude-stronger-source of gravity than all stars and galaxies in the Universe together*. This is the worst prediction in the history of physics which manifests that we miss something crucial and that our understanding of gravity of ordinary matter cannot be directly extended to the quantum vacuum.

No one can answer the question what the real physical solution to the cosmological constant problem is. However, we can ask a different question, *what is the simplest (not necessarily physical) solution that we can imagine*? The simplest imaginable solution (apart from physics, from the logical and mathematical point of view) is to assume that *quantum vacuum fluctuations are virtual gravitational dipoles*. If quantum vacuum fluctuations are gravitational dipoles, the total gravitational charge of the quantum vacuum in the Universe is trivially equal to zero. While this should be self-evident, just to be sure that everything is clear, let us note that in an analogous way, the total *electric charge* of the quantum vacuum in the Universe is zero, because particle-antiparticle pairs of charged particles are always electric dipoles.

If the total gravitational charge of the quantum vacuum in the Universe is zero, the quantum vacuum *with its natural random orientation of gravitational dipoles is not a source of gravity*. However, such a quantum vacuum can become a source of gravity through the mechanism of the gravitational polarisation (and the corresponding effective gravitational charge density) caused by the immersed Standard model matter.

The intriguing surprise of the above consideration is that the effective gravitational charge density *is of the right magnitude* to explain phenomena usually attributed to dark matter and dark energy. What to say. I am a theoretical physicist and I know well how strong and difficult to address are the mainstream arguments against the hypothesis that quantum vacuum fluctuations can be virtual gravitational dipoles. However we live in the time of perhaps the greatest crises in the history of physics and we cannot simply neglect possibilities saying that it is impossible.

Of course it is quite possible that the hypothesis that quantum vacuum fluctuations are gravitational dipoles is wrong, but even if wrong it is important as an invitation for thinking and a showcase of different scenarios we can imagine and must imagine in order to eventually establish the quantum vacuum as a source of gravity. In addition, if the hypothesis is wrong there are so many coincidences (see for example Hajdukovic 2010, 2013) suggesting that something different, but to some extent similar to our theory, must exist.

Quantum vacuum as a source of gravity is incompatible with two competing paradigms, hypothetical dark matter and hypothetical modification of gravity. For instance, if eventually dark matter is discovered and well describes dynamics of a galaxy, addition of the quantum vacuum as a new source of gravity will perturb the achieved harmony between observations and theory. Concerning modification of gravity, the impact of a major new source of gravity may be eventually fitted by an artificial, mathematical modification of gravity without a real physical modification of gravity.

Already for a few decades, thousands of scientists (it is a necessary collective effort in our struggle to reveal the secrets of Nature) consider hypothetical dark matter and hypothetical modification of gravity, while only a single person with important but limited help of S. Walter (Hajdukovic&Walter 2021) works on a much later proposed gravitational polarization of the quantum vacuum. Such disproportional efforts can be partially compensated for, when possible, the quantum vacuum is included as an alternative in attempt to explain observations. For instance, quantum vacuum should be included in studies of wide binary stars (Chae 2023, Hernandez 2023), but it was even not mentioned as a possibility.

In our opinion it is very likely that in the near future we will see the growing tensions between the Standard $\Lambda CDM$ cosmology and very sophisticated and precise astronomical observations. Tensions will trigger more serious consideration of non-mainstream alternatives, and hopefully, in the way presented here, or in a partially different way, the quantum vacuum will become an established source of gravity.

### 6.1 Quantum vacuum and the destiny of the Universe

Let us end with an inspiring showcase of beauty and richness of at first sight hidden interconnections. Celestial bodies compete for the gravitational polarization of the quantum vacuum; for instance the halo of the polarized quantum vacuum around our Earth would be much larger if Earth was much farther from the Sun (because the region in which its gravitational field is dominant would be larger). The question is if in principle the halo of a point-like body can be infinite or there is a maximum possible size of halo of a body. It seems reasonable to assume that after some large distance from the body, the random orientation of dipoles will prevail, i.e. gravitational field will not be sufficiently strong to align dipoles. If so, in principle, each body has a maximum possible size of the halo and the corresponding effective gravitational charge.

The destiny of the Universe depends on if a maximum possible size exist for each halo. If so, with the expansion of the Universe, the size of halos of galaxies and clusters of galaxies will increase till all halos reach their maximum size. With further expansion of the Universe the size of halos and the corresponding effective gravitational charge of the Universe will remain constant. If quantum vacuum is considered as a cosmological fluid it means that there is a transition from a fluid with negative pressure (when there is increase of the effective gravitational charge) to a pressureless fluid (with a constant effective gravitational charge). Hence, the accelerated expansion of the Universe will end. The endless accelerated expansion might be possible if individual halos can have infinite size.

# Appendix A

## Two point like bodies and islands of the quantum vacuum with a negative effective gravitational charge

Let us focus on the case of two point-like bodies with masses $M$ and $m$ at mutual distance $D$. Because of cylindrical symmetry we will use the cylindrical coordinates (See Figure 1) usually denoted $(z, \rho, \varphi)$; however, in order to avoid confusion between density $\rho_{qv}$ and coordinate $\rho$, we will use the notation $(z, s, \varphi)$ and will denote the corresponding unit vectors $(\mathbf{e}_z, \mathbf{e}_s, \mathbf{e}_\varphi)$.

The Newtonian acceleration $\mathbf{g}_N = g_{Nz}\mathbf{e}_z + g_{Ns}\mathbf{e}_s$ can be easily deduced from Fig. A1,

$$\mathbf{g}_N = -\left\{\frac{GMz}{(z^2+s^2)^{3/2}} - \frac{Gm(D-z)}{[(D-z)^2+s^2]^{3/2}}\right\}\mathbf{e}_z - \left\{\frac{GMs}{(z^2+s^2)^{3/2}} + \frac{Gms}{[(D-z)^2+s^2]^{3/2}}\right\}\mathbf{e}_s \quad (A1)$$

The corresponding magnitude of the Newtonian acceleration is:

$$g_N = G\sqrt{\frac{M^2}{(z^2+s^2)^2} + \frac{2mM(z^2+s^2-Dz)}{(z^2+s^2)^{3/2}[(D-z)^2+s^2]^{3/2}} + \frac{m^2}{[(D-z)^2+s^2]^2}} \quad (A2)$$

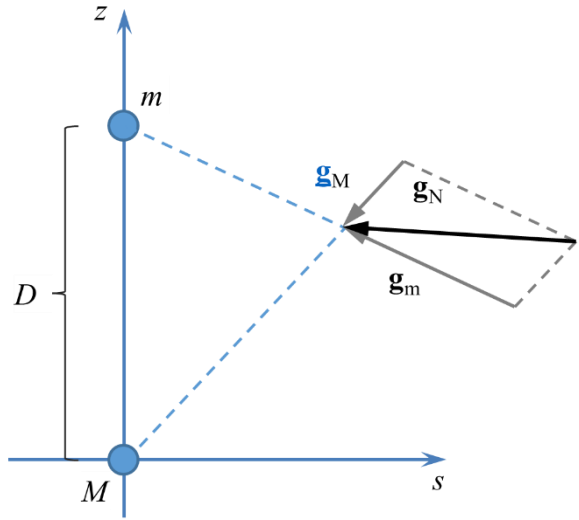

**Figure A1**. The resultant Newtonian gravitational field $\mathbf{g}_N$ of two point-like bodies with mass M and m at mutual distance $D$, has cylindrycal symmetry that trivially leads to Eq. (5).

According to equations (1) and (10), the effective gravitational charge density is

$$\rho_{qv}(z, s, \varphi, M, m, D) = P_{gmax}\boldsymbol{\nabla}\cdot\left[tanh\left(\sqrt{\frac{g_N}{4\pi G P_{gmax}}}\right)\frac{\mathbf{g}_N}{g_N}\right] \quad (A3)$$

After calculation of the divergence, the effective gravitational charge density is a messy *but explicit function* of cylindrical coordinates, masses $M$ and $m$, and the distance $D$ between bodies. Knowing distribution of the effective gravitational charge, the gravitational acceleration caused by quantum vacuum may be calculated.

In principle, Eq.(A3) can be applied to wide binary stars which can be well approximated by an ideal two body system; i.e. in binary systems in which the internal gravitational field is much stronger than the external one. Unfortunately, wide binary stars are very far; relatively low precision of measurements and especially very long orbital periods make impossible to test gravity in individual systems. However, it is possible to study statistically the internal kinematics of wide binary populations that can be selected from the Gaia DR3 database.

Let us apply Eq.(A3) to the Sun and a planet. At a distance of 1000AU from the Sun, the magnitude of acceleration caused by the Sun is about $5.9 \times 10^{-9} \, m/s^2$; hence two orders of magnitude larger than the acceleration $g_{qvmax} = 4\pi G P_{gmax}$. Consequently, with very high accuracy, hyperbolic tangent in Eq.(A3) is equal to 1; there is saturation of the polarized quantum vacuum in Solar System (non-saturation might exist only in relatively small regions in the vicinity of the point where gravitational fields of different bodies cancel each other).

In cylindrical coordinates, in the case of saturation, Eq.(A3) can be rewritten as:

$$\rho_{qv}(z,s,\varphi) = -P_{gmax}\left[\frac{\partial}{\partial z}\left(\frac{g_{Nz}}{g_N}\right) + \frac{1}{s}\frac{\partial}{\partial s}\left(s\frac{g_{Ns}}{g_N}\right)\right] \tag{A4}$$

Using equations (A1), (A2) and (A4), for *the region of saturation*, the effective gravitational charge density, induced by 2 bodies of masses M and m at mutual distance D, can be written as an explicit function of all variables:

$$\rho_{qv}(z,s,\varphi) =$$
$$P_{gmax}\left(\frac{G}{g_N}\right)^3 \left\{\frac{2M^3}{(z^2+s^2)^{7/2}} + \frac{2m^3}{[(D-z)^2+s^2]^{7/2}} + M^2 m \frac{6z^4+6s^4+12s^2z^2-16Dz^3-16Ds^2z+14D^2z^2+3D^2s^2-4D^3z}{(z^2+s^2)^3[(D-z)^2+s^2]^{5/2}} + Mm^2 \frac{6z^4+6s^4+12s^2z^2-8Dz^3-8Ds^2z+2D^2z^2-D^2s^2}{(s^2+z^2)^{5/2}[s^2+(D-z)^2]^3}\right\} \tag{A5}$$

Unfortunately, the explicit function given by Eq. (A5) is relatively long and complicated, so that main features are practically hidden; the use of numerical methods is inevitable. However, despite complexity, we can show analytically the most fundamental feature; in principle, in the region of saturation, and hence in the Solar System, the effective gravitational charge density can be negative.

Let us consider the case of two bodies with equal mass ($M = m$) at $z = D/2$. According to Eq. (A5), the corresponding density as a function of *s* is given by:

$$\rho_{qv}\left(z = \frac{D}{2}, s, \varphi\right) = \frac{2P_{pgmax}}{s}\frac{s^2 - D^2/8}{s^2 + D^2/4} \tag{A6}$$

The effective gravitational charge density determined by Eq. (A6) is negative for $s < D/\sqrt{8}$.

Now, let's have a look at Fig.A2a resulting from numerical calculations based on Eq. (A5). In the *zeroth approximation* (i.e., when the Sun is considered as the only source of gravity), there is a single, spherically symmetric halo of the polarised quantum vacuum, described by Eq.3; the effective gravitational charge density ($\rho_{qvSun} = 2P_{gmax}/r$) is always positive (we denoted it by green in figures). If a second body (in our case the planet Saturn) is included in calculations as a second source of gravitational polarization, this simple pattern of gravitational polarization is replaced by a more complex one. There are three main features of the new distribution of the effective gravitational charge density.

First, spherical symmetry converts to cylindrical symmetry; consequently, both the polarised quantum vacuum inside and outside the orbit have gravitational impact on Saturn.

Second, the effective gravitational charge density $\rho_{qv}(z,s)$, caused by the Sun and Saturn is different from the corresponding density $\rho_{qvSun}(z,s)$ calculated when the Sun was considered as the only source of polarization; as presented on Figure A2b, there are large regions (respectively denoted by blue and orange) in which $\rho_{qv}(z,s) > \rho_{qvSun}(z,s)$ and $\rho_{qv}(z,s) < \rho_{qvSun}(z,s)$.

Third, in a relatively large region around Saturn (with linear size of about one astronomical unit), the magnitude of the effective gravitational charge density is *much larger* than $2P_{gmax}/r$; and, in addition to positive there are also regions with a negative effective charge density (denoted by red colour). This red region is visible at the top of Fig.A2a. Hence there are "islands" (we may also say "clouds") of the quantum vacuum with negative effective gravitational charge.

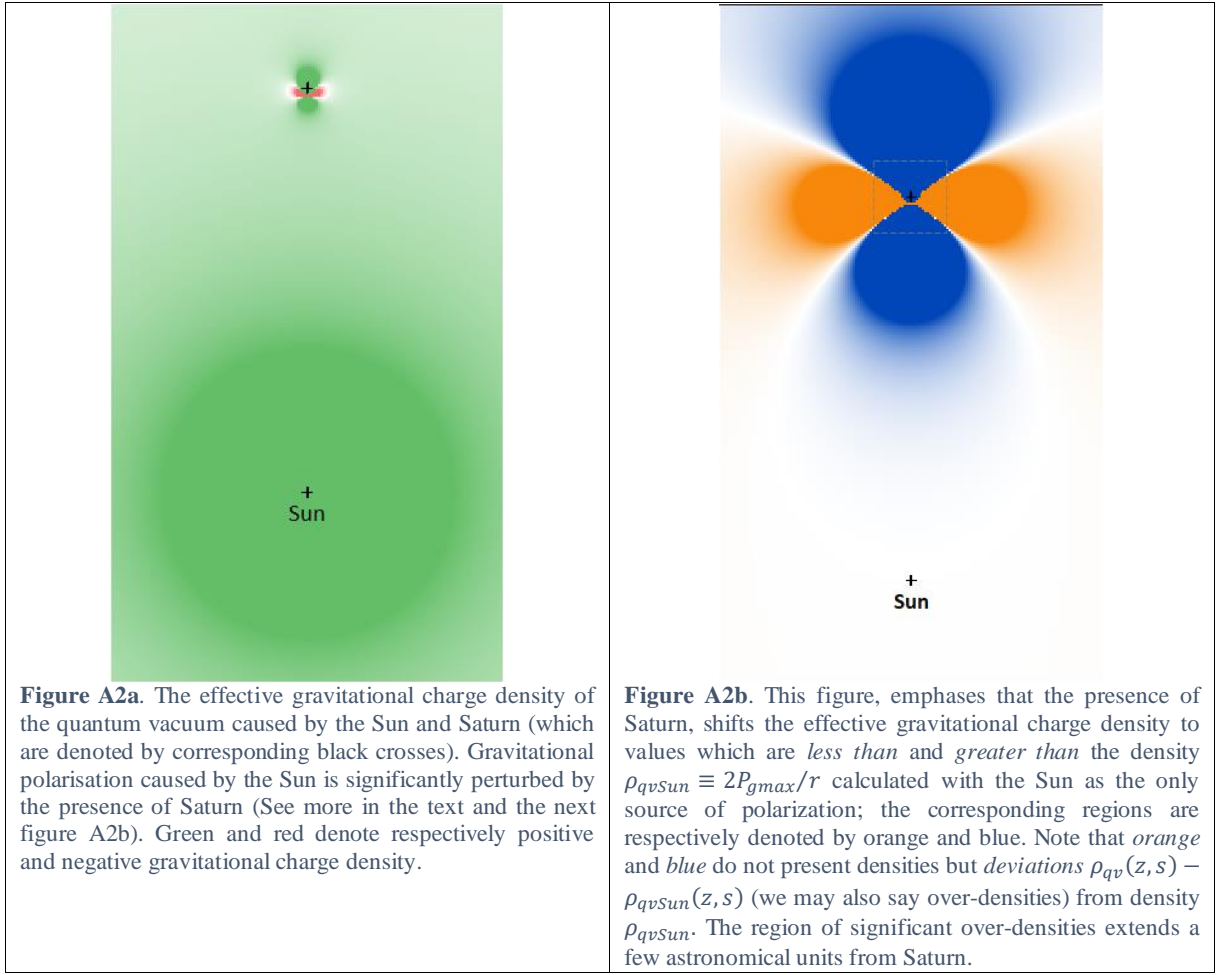

**Figure A2a**. The effective gravitational charge density of the quantum vacuum caused by the Sun and Saturn (which are denoted by corresponding black crosses). Gravitational polarisation caused by the Sun is significantly perturbed by the presence of Saturn (See more in the text and the next figure A2b). Green and red denote respectively positive and negative gravitational charge density.

**Figure A2b**. This figure, emphases that the presence of Saturn, shifts the effective gravitational charge density to values which are *less than* and *greater than* the density $\rho_{qvSun} \equiv 2P_{gmax}/r$ calculated with the Sun as the only source of polarization; the corresponding regions are respectively denoted by orange and blue. Note that *orange* and *blue* do not present densities but *deviations* $\rho_{qv}(z,s) - \rho_{qvSun}(z,s)$ (we may also say over-densities) from density $\rho_{qvSun}$. The region of significant over-densities extends a few astronomical units from Saturn.

The next figure A3 presents the red island of the quantum vacuum (with a negative effective gravitational charge density) which is the closest one to the Earth. In fact, the island is volume obtained by rotation of the red surface for $\pi$ around *z* axis (or rotation for $2\pi$ of the left or right half of the red surface).

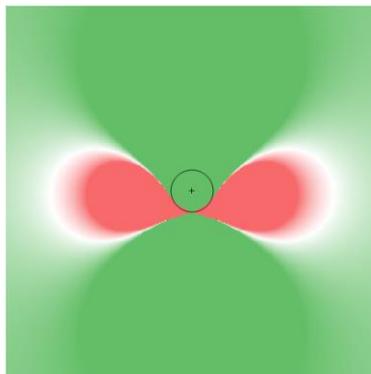

**Figure A3**. The island of the quantum vacuum with a negative effective gravitational charge (denoted by red) within the sea of positive charge (denoted by green). This is a typical form for all the planets but here we present calculations for the Earth and the Sun. The Earth is denoted by the black cross, while the Sun is far below and not visible on this plot. The radius of the circle around the Earth is 260,000 km, while the orbit of the Moon is 384,000 km and hence a part of the orbit of the Moon goes through the red region producing its perturbation.

The radius of the small circle around Earth is $r_c = D/(1 + \sqrt{M/m})$ where $M$ and $m$ are respectively the mass of the Sun and planet (in this case Earth) and D is the distance between them (For Earth $D = 1.5 \times 10^8 km$). Let us underscore that on the z axis at distance $z_c = D\sqrt{M/m}$ from the Sun and distance $r_c$ from the Earth, Newtonian accelerations caused by the Sun and the Earth cancel each other.

For Earth $r_c \approx 260\,000\ km$, while the linear size of the red surface is nearly four million kilometres (about $15 r_c$). Note that the radius of the Moon's orbit around Earth is about $1.5 r_c$; hence, the orbit of the Moon partially goes through the island. It remains a challenge to reveal if the predicted island (cloud) of the negatively charged quantum vacuum within the Earth-Moon system produces a measurable gravitational impact. In a similar way it remains a challenging question if the existence of such a cloud might be eventually revealed by careful study of the trajectory of a spacecraft launched from the Earth and approaching Jupiter or Saturn through the cloud.

Negative effective gravitational charges (denoted by $m_{qv}^-$) of islands corresponding to planets of the Solar System are given in Table A1. The volume element in cylindrical coordinates is $dV = sdzdsd\varphi$ and because of cylindrical symmetry results in Table 1 follow from numerical integration ($m_{qv}^- = 2\pi \iint \rho_{qv}(z,s) s\, dzds$). Islands corresponding to the four big planets have a relatively big mass close in magnitude to the mass of significant dwarf planets as for instance (55637) 2002 UX25 and 90482 Orcus. The total negative effective gravitational charge in the Solar System should be close in magnitude to the total mass of the main asteroid belt between Mars and Jupiter.

| Planet | $m_{qv}^-$ [kg] | $g_{qv}^-$ [$m/s^2$] |
|---|---|---|
| Mercury | $-3.21 \times 10^{14}$ | $-4.83 \times 10^{-12}$ |
| Venus | $-1.27 \times 10^{16}$ | $-4.80 \times 10^{-12}$ |
| Earth | $-2.94 \times 10^{16}$ | $-4.80 \times 10^{-12}$ |
| Mars | $-9.12 \times 10^{15}$ | $-4.83 \times 10^{-12}$ |
| Jupiter | $-9.62 \times 10^{19}$ | $-4.18 \times 10^{-12}$ |
| Saturn | $-1.29 \times 10^{20}$ | $-4.45 \times 10^{-12}$ |
| Uranus | $-1.11 \times 10^{20}$ | $-4.68 \times 10^{-12}$ |
| Neptune | $-3.12 \times 10^{20}$ | $-4.66 \times 10^{-12}$ |

**Table A1**. Second column gives negative effective gravitational charge $m_{qv}^-$ of islands corresponding to planets. As an illustration of the size of the gravitational impact of these islands, the third column gives the acceleration caused by an island at the centre of the planet.

The simplest way to get a rough idea of the gravitational impact of a negatively charged island is to calculate acceleration $g_{qv}^-$ caused by that island at points on the z axis close to the island. Acceleration at a point ($z = z_0, s = 0$) is given by integration over the volume of the island:

$$g_{qv}^- = 2\pi G \iint \rho_{qv}(z,s) \frac{(z-z_0)s}{[(z-z_0)^2+s^2]^{3/2}} dzds \tag{A7}$$

Equation (A7) is a consequence of the fact that to each surface element $dzds$ corresponds to a ring of volume $dV = 2\pi s\, dzds$ with the effective gravitational charge $dm_{qv}^- = 2\pi \rho_{qv}(z,s)s\, dzds$. The acceleration caused by this elementary ring is $dg_{gv}^- = G(z - z_0)dm_{qv}^-/[(z - z_0)^2 + s^2]^{3/2}$ leading to Eq. (A7). As an example, the numerical value for acceleration at point $z = D$ (i.e., at the centre of the planet) is given for all islands in the third column of Table 1.

Figure A2a and numerical values for acceleration in Table A1 raise the question if the gravitational impact of the quantum vacuum causes a very tiny violation of the Weak Equivalence Principle (i.e., universality of free fall) that is a cornerstone of both, Newtonian and general relativistic gravity. Imagine that in Figure A2a, instead of Saturn there is a planet of a different mass.

As intuitively expected, (and confirmed by Eq. (A7)) a different mass leads to a different distribution of the effective gravitational charge density; hence it seems likely that the gravitational impact of the quantum vacuum on a body depends on its mass $m$.

In brief, we revealed an unprecedented possibility that the Solar System contains islands of the quantum vacuum with a negative effective gravitational charge density. Theoretical physics is in crisis, perhaps in the greatest crisis in its history (let us mention only the unsolved mysteries of phenomena usually attributed to hypothetical dark matter and dark energy and not the less mysterious asymmetry between matter and antimatter in the Universe); hence, we must stay open-minded to different scenarios even if they look very unlikely from the point of view of our current knowledge. Hopefully this paper would stimulate astronomers and physicists for further considerations.